\documentclass[11pt]{article}

\usepackage{amsthm}
\usepackage{amsmath}
\usepackage{thmtools,thm-restate}
\usepackage{amssymb}
\usepackage{graphicx}
\usepackage[margin=2cm, includefoot]{geometry}
\usepackage{commath}
\usepackage{xcolor}
\usepackage{color}
\usepackage{url}
\usepackage{algorithm}
\usepackage{algpseudocode}
\usepackage{bm}
\usepackage{verbatim}
\usepackage{subcaption}
\usepackage{enumitem}
\usepackage[toc,page]{appendix}

\newcommand{\I}{\iota}
\newcommand{\R}{\mathbb{R}}
\newcommand{\Q}{\mathbb{Q}}
\newcommand{\C}{\mathbb{C}}
\newcommand{\K}{\mathbb{K}}
\newcommand{\Z}{\mathbb{Z}}

\newcommand{\hx}{\hat{x}}
\newcommand{\E}{\mathbb{E}}

\newtheorem{thm}{Theorem}
\numberwithin{equation}{section}
\numberwithin{thm}{section}

\newtheorem{conj}[thm]{Conjecture}
\newtheorem{cor}[thm]{Corollary}

\newtheorem{remark}[thm]{Remark}
\usepackage{authblk}

\begin{document}
\title{Signal recovery from  a few linear measurements of its high-order spectra}
\author{Tamir Bendory, Dan Edidin, and Shay Kreymer}

\maketitle
\begin{abstract}
The $q$-th order spectrum is a  polynomial of degree $q$ in the entries of a signal $x\in\C^N$, which is invariant under circular shifts of the signal. For $q\geq 3$, this polynomial  determines  the signal uniquely, up to a circular shift, and is called a high-order spectrum.
The high-order spectra, and in particular the bispectrum ($q=3$) and the trispectrum ($q=4$), play a prominent role in various statistical signal processing and imaging applications, such as phase retrieval and single-particle reconstruction.
However, the dimension of the  $q$-th order spectrum is $N^{q-1}$, far exceeding the dimension of~$x$,
 leading to increased computational load and storage requirements. In this work, we show that it is unnecessary to store and process the full high-order spectra: a signal can be uniquely characterized up to {symmetries},
 from only $N+1$ linear measurements of its high-order spectra.
The proof relies on tools from algebraic geometry and is corroborated by numerical experiments.
\end{abstract}


\section{Introduction}
Let $\hx\in\C^N$ be the discrete Fourier transform (DFT) of a signal $x\in\C^N$.
The \emph{bispectrum} of $x$ is defined by  the triple products
\begin{equation} \label{eq:bispectrum}
	M_3(x)[k_1,k_2] := \hx[k_1]\hx[k_2]\hx[-k_1-k_2], \qquad  k_1,k_2=0,\ldots,N-1,
\end{equation}
where all indices should be considered as modulo $N$.
The bispectrum is designed to be invariant under circular shifts, namely,  under the mapping $x[n]\mapsto x[n-s]$  for any $s\in\Z$.
This is true since circularly shifting $x$ by $s$ entries is equivalent to
multiplying its $k$-th DFT coefficient by the phase~$e^{-2\pi\I k s/N}$,{ where $\I=\sqrt{-1}$.}
In particular, denoting the shifted signal by $x_s$, it is easy to see that
\begin{equation}
	\begin{split}
	M_3(x_s)[k_1,k_2] &:= \hx_s[k_1]\hx_s[k_2]\hx_s[-k_1-k_2] \\ & = \hx[k_1]e^{-2\pi\I k_1 s/N}\hx[k_2]e^{-2\pi\I k_2 s/N}\hx[-k_1-k_2]e^{2\pi\I (k_1+k_2) s/N} \\ &= M_3(x)[k_1,k_2],
	\end{split}
\end{equation}
for any $s,k_1,k_2=0,\ldots,N-1$.
Since each entry of the bispectrum is a monomial of degree 3 it is also invariant under multiplication by  $e^{2\pi\I \ell/3}$ for $\ell=0,1,2$ and we refer to
it as a third-order invariant.
In addition, the bispectrum determines almost all signals uniquely, up to {symmetries}
(see, for example,~\cite{sadler1992shift,bendory2017bispectrum}).
The signal and all its {symmetries}
are called the \emph{orbit} of $x$, and thus we say that the bispectrum determines the orbit of $x$ uniquely, for almost any~$x$.

Similarly to the bispectrum,  the \emph{trispectrum}, a fourth-order invariant, is defined as
\begin{equation} \label{eq:trispectrum}
	M_4(x)[k_1,k_2,k_3] = \hx[k_1]\hx[k_2]\hx[k_3]\hx[-k_1-k_2-k_3], \qquad  k_1,k_2,k_3=0,\ldots,N-1.
\end{equation}
The trispectrum
enjoys  {similar}
properties
as the bispectrum: it is invariant under circular shifts {and multiplication by $e^{2\pi\I \ell/4}$ for $\ell=0,1,2,3$}, and determines almost all orbits uniquely.
The bispectrum and the trispectrum are known in the signal processing and statistics communities for many years~\cite{tukey1953spectral}, and have been used in
a variety of signal processing applications, such as separating Gaussian and non-Gaussian processes~\cite{brockett1988bispectral,bartolo2004non},    cosmology~\cite{luo1993angular,wang2000cosmic,hu2001angular,byrnes2006primordial}, seismic signal processing~\cite{matsuoka1984phase}, image
deblurring~\cite{chang1991blur}, feature extraction for radar~\cite{chen2008feature}, analysis
of EEG signals~\cite{ning1989bispectral}, classification~\cite{zhao2014rotationally},
and multi-reference alignment~\cite{bendory2017bispectrum,aizenbud2019rank,landa2019multi,bendory2021super}.

The bispectrum and the trispectrum can be further generalized to higher-order invariants.
 In particular, the $q$-th order invariant is defined by a product of $q$ DFT coefficients
\begin{equation} \label{eq:high-order-spectra}
	M_q(x)[k_1,\ldots,k_{q-1}]=\hx[k_1]\hx[k_2]\ldots\hx[k_{q-1}]\hx[-k_1-k_2,\ldots -k_{q-1}].
\end{equation}
For any $q\geq 3$,  $M_q(x)$ determines almost any orbit.
Throughout the work, we treat $M_q(x)$ as a column vector in $\C^{N^{q-1}}$, where the bispectrum corresponds to
$q=3$ and the  trispectrum to $q=4$.
We refer to~\eqref{eq:high-order-spectra} for $q\geq 3$ as \emph{high-order spectra}.
For $q=1$ and $q=2$,~\eqref{eq:high-order-spectra} reduces to, respectively, the mean and the power spectrum which do not determine a signal or its orbit uniquely~\cite{bendory2017fourier} {unless additional information on the signal, such as sparsity, is available~\cite{bendory2020toward,ghosh2021multi}.}.

This work is motivated by two imaging applications: phase retrieval and single-particle reconstruction. These are introduced in detail in Section~\ref{sec:motivation}.
In the former application, linear measurements of the trispectrum naturally arise in the data generative model, while the bispectrum was exploited in the latter to design computationally efficient algorithms.
However, the high-dimensionality of  high-order spectra raises a challenge: a  high-order spectrum consists of $N^{q-1}$ entries and thus its dimension far exceeds the signal's dimension, especially for large $N$.
This naturally raises the question whether the full high-order spectrum is required for signal recovery, or whether its  concise summary suffices.
To answer this question, we consider the problem of recovering a signal from  linear measurements of its high-order spectrum.
 Specifically, the measurement model reads
\begin{equation} \label{eq:measurement}
	y = AM_q(x),
\end{equation}
where $A\in\C^{K\times N^{q-1}}$ and $M_q(x)\in\C^{N^{q-1}}$ so that $y\in\C^{K}$.
Trivially, if $K=N^{q-1}$ and $A$ is invertible, then if the orbit of $x$ can be recovered from $M_q(x)$ (which is true for almost all signals for $q\geq 3$), it can  be also recovered from~$A^{-1}y$.
However, this work establishes that the invertibility of~$A$ is not a necessary condition.
Our main result shows that the orbit of a  signal can be  determined uniquely from $y$
even if  the rank of $A$ is as low as~$N+1$.
In other words, only $N+1$ (generic) linear measurements of a high-order spectrum suffice to determine a (generic) signal uniquely, up to {symmetries.} 
This result is summarized by the following theorem.
\begin{thm}\label{thm.informal}
{Suppose that the orbits of generic signals $x\in\C^N$ are determined uniquely from the $q$-th order spectrum~\eqref{eq:high-order-spectra}. Then, the orbit of a generic signal $x$ is also determined uniquely from~\eqref{eq:measurement} for almost any matrix $A\in\C^{K\times N^{q-1}}$ with $K\geq N+1$.}
\end{thm}


Theorem~\ref{thm.informal} is a corollary of a more general theoretical result---Theorem~\ref{thm.formal}---which is based on algebraic geometry tools.
While the result holds for any high-order spectra $q\geq 3$, our main interest is the bispectrum ($q=3$) and the trispectrum ($q=4$). Thus, we state the following  corollary.


\begin{cor}
	Almost every signal $x\in\C^N$ is determined uniquely, up to symmetries, from generic $N+1$ linear measurements of its bispectrum or its trispectrum.
\end{cor}

Section~\ref{sec:numerics} corroborates our theoretical results with a numerical study.  We show that indeed a signal can be recovered, up to {symmetries,}
from slightly more than $N$ measurements.
We also show numerically that  a signal can be recovered from a few of its high-order spectrum entries.
While our proof does not cover the latter case, in Section~\ref{sec:future} we formulate a conjecture stating that, with high probability, a signal can be recovered from ${O}(N)$ random samples of its high-order spectra.


\section{Motivation} \label{sec:motivation}

\subsection{Ultra-short pulse characterization using multi-mode fibers}

Femtosecond-scale pulses are a key ingredient in
investigations of ultrafast phenomena, such as chemical reactions, and electron dynamics in atoms and molecules~\cite{hu2006attosecond,demtroder2013laser,sheetz2009ultrafast,trebino2000frequency}.
In particular, characterizing the shape of an ultrashort optical pulse is an essential task.
Unfortunately, sensor technology does not yet have short enough response time to recover ultrashort pulses directly, and thus developing technological and computational methods to circumvent this barrier is required.
 For example, in a popular
method called frequency-resolved optical gating,
the sought signal interacts with shifted versions of itself, resulting in a quartic map that can be used to recover a signal (up to some intrinsic symmetries)~\cite{trebino2000frequency,bendory2020signal,bendory2017uniqueness,bendory2019blind}.

Recently, a novel method for pulse characterization  in a single-shot using multi-mode fibers (which are typically used for communication purposes) has been proposed and implemented~\cite{xiong2019deep,ziv2020deep}.
The technique uses a nonlinear  measurement of
transmitted light through a multi-mode fiber to extract the spectral phase of an optical pulse of interest. Two-photon absorption on an array of detectors produces a nonlinear pattern, from which the signal can be recovered.
This experimental technique has a number of advantages: it is a single-shot method, its experimental setup is very simple, and it produces accurate estimates  in the presence of noise (namely, it is robust against noise).
While this method
shows great potential and has attracted the attention of leading figures in the
optical imaging community, its mathematical foundations remain obscure.  Consequently, there is a great need to develop a supporting mathematical theory
that will allow the harnessing of the full potential of this  technique.

The problem of pulse characterization using multi-mode fibers can be mathematically formulated  as acquiring linear measurements of the pulse's trispectrum~\eqref{eq:trispectrum}~\cite{xiong2019deep}.
{In particular, each linear measurement (namely, each row of the matrix $A$~\eqref{eq:measurement}) is itself a trispectrum of a signal in $\C^N$.}
The results of this paper {suggest} that  one can acquire only a few samples (i.e., a few linear measurements of the pulse's trispectrum)
and still guarantee full recovery of the sought pulse.

\subsection{Invariants for reconstructing  molecular structures}
\label{sec:cryoEM}
Single-particle cryo-electron microscopy (cryo-EM) is an emerging technology
to reconstruct the high-resolution three-dimensional structure of macromolecules, such as proteins and viruses~\cite{bai2015cryo,nogales2015cryo,vinothkumar2016single}.
Recent substantial developments in the field has led to
an abundance of new molecular structures, garnering its recognition by the 2017 Nobel Prize in Chemistry.

In a cryo-EM experiment, biological macromolecules suspended
in a liquid solution are rapidly frozen into a thin ice
layer. The three-dimensional orientation of particles within the
ice are random and unknown. An electron beam then passes
through the sample, and a two-dimensional tomographic projection, called a
micrograph, is recorded. The goal is to reconstruct a high-resolution
estimate of the three-dimensional electrostatic potential of the molecule from a set of micrographs.
Under some simplifying assumptions, the cryo-EM problem entails estimating the  three-dimensional structure $X$ from multiple observations:
\begin{equation} \label{eq:cryo_em}
	I_i = PR_{\omega_i} X + \varepsilon_i, \qquad i=1,\ldots,N,
\end{equation}
where $P$ is a fixed tomographic projection, $R_{\omega_1},\ldots,R_{\omega_N}$ are random  three-dimensional rotations (elements of the group $SO(3)$), and $\varepsilon$ is a noise term.    The full mathematical model is elaborated in~\cite{bendory2020single}.

The main computational challenge in cryo-EM stems from the compounding effect of unknowing the three-dimensional rotations and the high noise level: the power of the noise might be~100 times greater than the power of the signal. Such noise levels hamper accurate estimation of the missing three-dimensional rotations~\cite{bendory2019multi,aguerrebere2016fundamental}. Therefore, it is common to estimate the three-dimensional structure directly, without estimating the missing rotations, for example, by maximizing the marginal likelihood~\cite{scheres2012relion}.

Zvi Kam was the first to propose  circumventing rotation estimation by computing polynomials of the signal  that are invariant to  three-dimensional rotations~\cite{kam1980reconstruction}.
Those polynomials can be understood as an extension of the one-dimensional bispectrum~\eqref{eq:bispectrum} to the statistical model of cryo-EM~\eqref{eq:cryo_em}\footnote{We refer the reader to~\cite{kakarala2009completeness} for a rigorous extension of the concept of bispectrum  to any compact group.}.
Kam's idea was extended in recent years and used to construct ab inito models, see for example~\cite{levin20183d,bendory2018toward,sharon2020method,lan2021random}.
In addition,
it was thoroughly studied for multi-reference alignment and multi-target detection: mathematical abstractions of the cryo-EM problem~\cite{bendory2017bispectrum,boumal2018heterogeneous,perry2019sample,ma2019heterogeneous,bandeira2017estimation,bendory2019multi,lan2020multi,bendory2021multi,kreymer2021two}; see further discussion on the multi-reference alignment model in Section~\ref{sec:future} and Conjecture~\ref{conj:mra}.
The invariants-based approach was also studied for the problem of X-ray free-electron lasers (XFEL): a new exciting technology for single-particle reconstruction~\cite{maia2016trickle,starodub2012single,kurta2017correlations}.

One of the main challenges to apply Kam's method  to experimental cryo-EM datasets is that computing the bispectrum (or higher-order spectra) inflates the dimensionality of the problem.
For example, for a three-dimensional structure of size $L\times L\times L$,
the bispectrum proposed by Kam~\cite{kam1980reconstruction} {is composed of} $O(L^5)$ elements, namely, it increases the dimensionality by a factor of  $O(L^2)$.  Our results indicate that one can safely reduce the bispectrum's dimension by multiplying  it by a random matrix {of significantly lower rank},  without losing information.

\section{Theory}
\label{sec:theory}

The goal of this section is to state and prove the main  theoretical contribution of this paper, which implies
Theorem~\ref{thm.informal} as a corollary.
Appendix~\ref{sec:appendix} surveys  some definitions and results from the field of algebraic geometry required to fully comprehend the result.

\subsection{Main theoretical result}
We start by formulating our problem in algebraic geometry terms.
  Let $G$ be a finite group acting on $\C^N$ and suppose
  that we are given $R$ polynomial functions
$f_1, \ldots , f_R,$ which are invariant under the action of $G$.
These functions define a polynomial map $T \colon \C^N \to \C^R$.
Because the functions are $G$ invariant, we obtain
a map of varieties $\tilde{T} \colon \C^N/G \to \C^{R}$,
where $\C^N/G$ is the variety of $G$ orbits in $\C^N$.
Now, let
$A \in \C^{K \times R}$ be a matrix and consider
the composite map $S = A \circ \tilde{T} \colon \C^N/G \to \C^K$.

We are now ready to state the main theorem of this paper.
\begin{thm} \label{thm.formal}
  If the map $\tilde{T} \colon \C^N/G \to \C^R$  is
  birational onto its image, then for generic choice of
  matrix $A$ of rank at least $N+1$, the map $S= A \circ \tilde{T}$ is also birational
  onto its image, meaning that the generic orbit $x \in \C^N/G$ can be recovered from the measurements $A \circ \tilde{T}$.
\end{thm}

  In less precise terms, Theorem~\ref{thm.formal} states that if the measurements
  determined by the map
  $\tilde{T}$ are sufficient to recover generic $G$ orbits, then for a generic choice of {a}
  $K \times R$ matrix of rank at least $N+1$,  the measurements $A \circ \tilde{T}$
    are also sufficient.

  \begin{remark} \label{rem.ref1comment3}
    Although  Theorem \ref{thm.formal} is stated for complex signals (i.e., vectors in $\C^N$) the field $\C$ can be replaced by other fields such as the reals $\R$ or even the rationals $\Q$.
    \end{remark}
\subsection{Proof of Theorem~\ref{thm.formal}}

  Let $X$ be the closure of the image of $\C^N/G$ in $\C^R$ under the map
  $\tilde{T}$.
  Since $\tilde{T}$
  maps $\C^N/G$ birationally onto its image, $X$ is an $N$-dimensional subvariety of $\C^R$. We must show that for a generic matrix $A$
  of rank at least $N+1$, $X$ maps birationally onto its image under the
  linear transformation $\C^R \stackrel{A} \to \C^K$.

 A fundamental theorem in algebraic geometry \cite[Theorem 1.8]{shafarevich2013basic} states that any $N$-dimensional affine variety admits
  a birational map to a hypersurface in $\C^{N+1}$.
  However, the proof (cf.~\cite[Proposition A.7]{shafarevich2013basic})
  of this result shows that if we consider a generic (hence linearly independent) collection of
  $N+1$ linear forms $l_1,\ldots , l_{N+1}$, then the projection
  $\C^{R} \to \C^{N+1}$, $x \mapsto (l_1(x), \ldots ,l_{N+1}(x))$
maps~$X$ birationally onto its image.
  Now if $A$ is a generic $K \times R$ matrix of rank at least $N+1$,
  then the first $N+1$ rows of $A$ define a generic collection
  of $N+1$ linear forms. Hence, the composite $\C^R \stackrel{A} \to \C^{K}
  \stackrel{\pi_{N+1}} \to \C^{N+1}$ is as in the proof of~\cite[Theorem 1.8]{shafarevich2013basic}, where $\pi_{N+1}$ is the projection onto the first $N+1$ coordinates.
 Therefore, if the map $X \to (\pi_{N+1} \circ A)(X)$ is birational, so the map $X \to A(X)$
 must also be birational.

\begin{remark}
    The proof of Theorem \ref{thm.informal} follows by taking $T$ to be the $q$-th order spectrum with $q \geq 3$ and in our setup $R=N^{q-1}$. The finite group
    $G$ is the product group $\Z_N \times \Z_{q}$. The $\Z_N$ factor acts in the time domain by cyclic shifts and the $\Z_{q}$ factor is identified with the group
    $q$-th roots of unity acting by scalar multiplication. 
  \end{remark}

 \subsection{Example}
  While Theorem~\ref{thm.informal} implies that $N+1$ generic linear measurements of the $q$-th {order} spectrum are sufficient to recover generic signals, it does not indicate which linear measurements suffice. The following heuristic shows
  that we can recover a large class of real signals $x\in\R^N$, up to circular shifts, from a
  small set of bispectrum or trispectrum measurements~\cite{bendory2017bispectrum}.
	  To this end, we make three
	assumptions. First, the power spectrum of the signal does not vanish
	and is known, and thus we can assume that the magnitudes of all Fourier
	coefficients are one; this is indeed the case
	in ultra-short pulse characterization using multi-mode fibers~\cite{xiong2019deep}. In addition, we assume that the mean of the
	signal is known, and thus~$\hx[0]$ is known (and real). Finally, we assume that
	the phase of $\hx[1]$ is an $N$-th root of unity. Recall that the circular
	shift symmetry implies that $\hx[1]$ can be multiplied by~$e^{2\pi\I
		m/N}$ for an arbitrary $m\in\Z$. Therefore, the third assumption
	implies that we can fix $\hx[1]=1$ without loss of generality. Based on
	these three assumptions, one can easily read off $\hx[2]$ from
	$M_3(x)[2,N-1] :=
	\hx[2]\overline{\hat{x}}[1]\overline{\hat{x}}[1],$ where
	$\overline{\hat{x}}[\ell]$ is the conjugate of ${\hat{x}}[\ell]$ and we used the symmetry ${\hat{x}}[\ell]=\overline{\hat{x}}[-\ell]$
 since $x$ is real.
	Continuing recursively, the $k$-th Fourier coefficient can be
	determined, given $\hx[0],\ldots,\hx[k-1]$, from $M_3(x)[k,N-1] :=
	\hx[k]\overline{\hat{x}}[1]\overline{\hat{x}}[k-1].$
	The same recursion can be applied to the trispectrum.

\section{Numerical experiments}
\label{sec:numerics}

We conducted {three} sets of numerical experiments.
The first experiment examines recovering a signal~$x\in\R^N$
 from $K$ random linear  measurements of its bispectrum and trispectrum.
   Recall that we treat the bispectrum and the trispectrum as vectors in {$\C^{N^2}$ and $\C^{N^3}$}, respectively, and that we observe
 \begin{equation} \label{eq:meas_numerical_exp}
 	y = AM_q(x), \quad {q=3,4}.
 \end{equation}
 In the second experiment,  each row of the matrix $A$   is a bispectrum (for $q=3$) or a trispectrum ($q=4$) of a random signal. Therefore, the measurement matrix is more structured than in the first experiment. This experiment simulates  the setup of ultra-short pulse characterization using multi-mode fibers---one of the motivating applications of this paper (see Section~\ref{sec:motivation}).
 The {third} experiment studies signal recovery from $K$ random samples of its bispectrum and trispectrum. 
The sampling problem can be formulated as in~\eqref{eq:meas_numerical_exp}, where the matrix {is} $A\in\{0,1\}^{K \times N^{q-1}}$, each row of $A$ consists of only one  non-zero entry, and each column has at most one non-zero entry.
{We note that the second and third  cases are not covered by Theorems~\ref{thm.informal} and \ref{thm.formal} (see further discussion in Section~\ref{sec:future}).}
 In {all} experiments, the {entries of }
  $x\in\R^N$  were drawn independently from a normal distribution with zero mean and variance one. 
  We note that   the symmetry groups of real signals are smaller since rescaling by a root of unity can only be a sign flip and only if $q$ is even.
  For the bispectrum, we used signals of length $N = 30$, and for the trispectrum $N = 10$.

To recover the signal, we formulated a non-convex least squares problem
\begin{equation}
	\label{eq:ls}
	\min_{x\in\R^N} \|y - A M_q(x)\|_2^2,
\end{equation}
which was minimized using a standard steepest decent algorithm. 
To account for the non-convexity of the problem, we initialized the algorithm from three  random {points}, resulting in three candidate solutions. The candidate solution that attained the smallest value of~\eqref{eq:ls} was declared as the signal estimate. From our experiments, it seems that three initializations suffice to avoid getting trapped in a local minimum. This observation concurs with previous papers on bispectrum inversion, indicating that the non-convexity of bispectrum inversion is often times benign~\cite{bendory2017bispectrum,boumal2018heterogeneous}.

 To account for the circular shift symmetry, the {bispectrum} recovery error  is computed by
 \begin{equation}
 	\text{{bispectrum} relative error} = \min_{s=0,\ldots,N-1} \frac{\|R_s\hat{x} - x\|_2}{\|x\|_2},
 \end{equation}
 where $\hat{x}$ is the signal estimate, and $R_s$ is the operator that circularly shifts the signal by $s$ entries.
 Similarly, to account for the additional sign flip symmetry,
 the trispectrum recovery error  is defined as
 \begin{equation}
 	\text{trispectrum relative error} = \min_{s=0,\ldots,N-1, z=\pm 1} \frac{\|z\cdot R_s\hat{x} - x\|_2}{\|x\|_2}.
 \end{equation}
 A trial was declared successful if the relative recovery error dropped below {$5\times 10^{-5}$}, {and all  figures show the success rate over 1000 trials.}
 The code to reproduce all experiments is publicly available at
{ \url{https://github.com/krshay/recovery-high-order-spectra}.}

\begin{figure}
	\begin{subfigure}[ht]{0.45\columnwidth}
		\centering
		\includegraphics[width=\columnwidth]{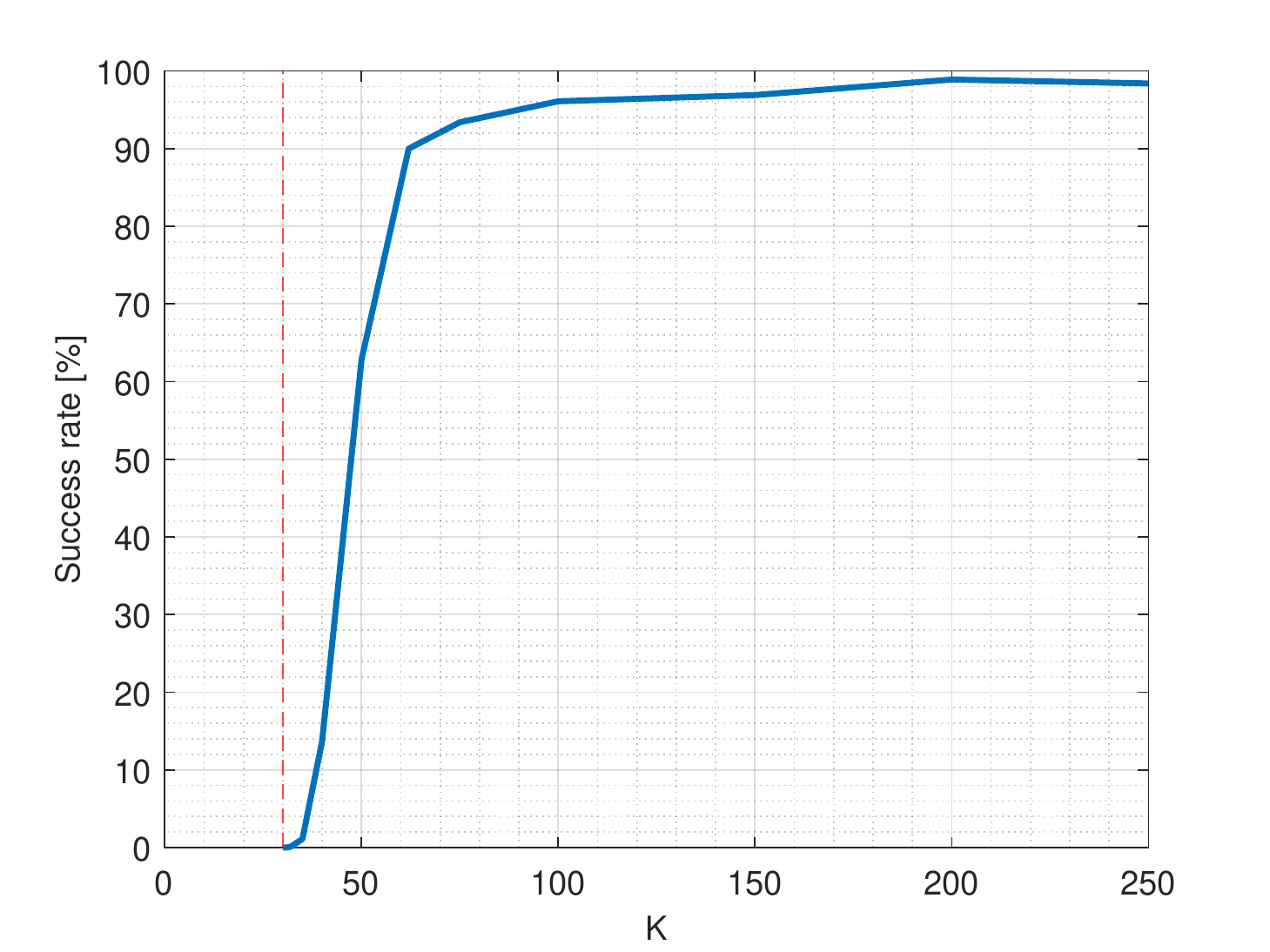}
		\caption{Bispectrum, $N=30$}
	\end{subfigure}
	\hfill
	\begin{subfigure}[ht]{0.45\columnwidth}
		\centering
		\includegraphics[width=\columnwidth]{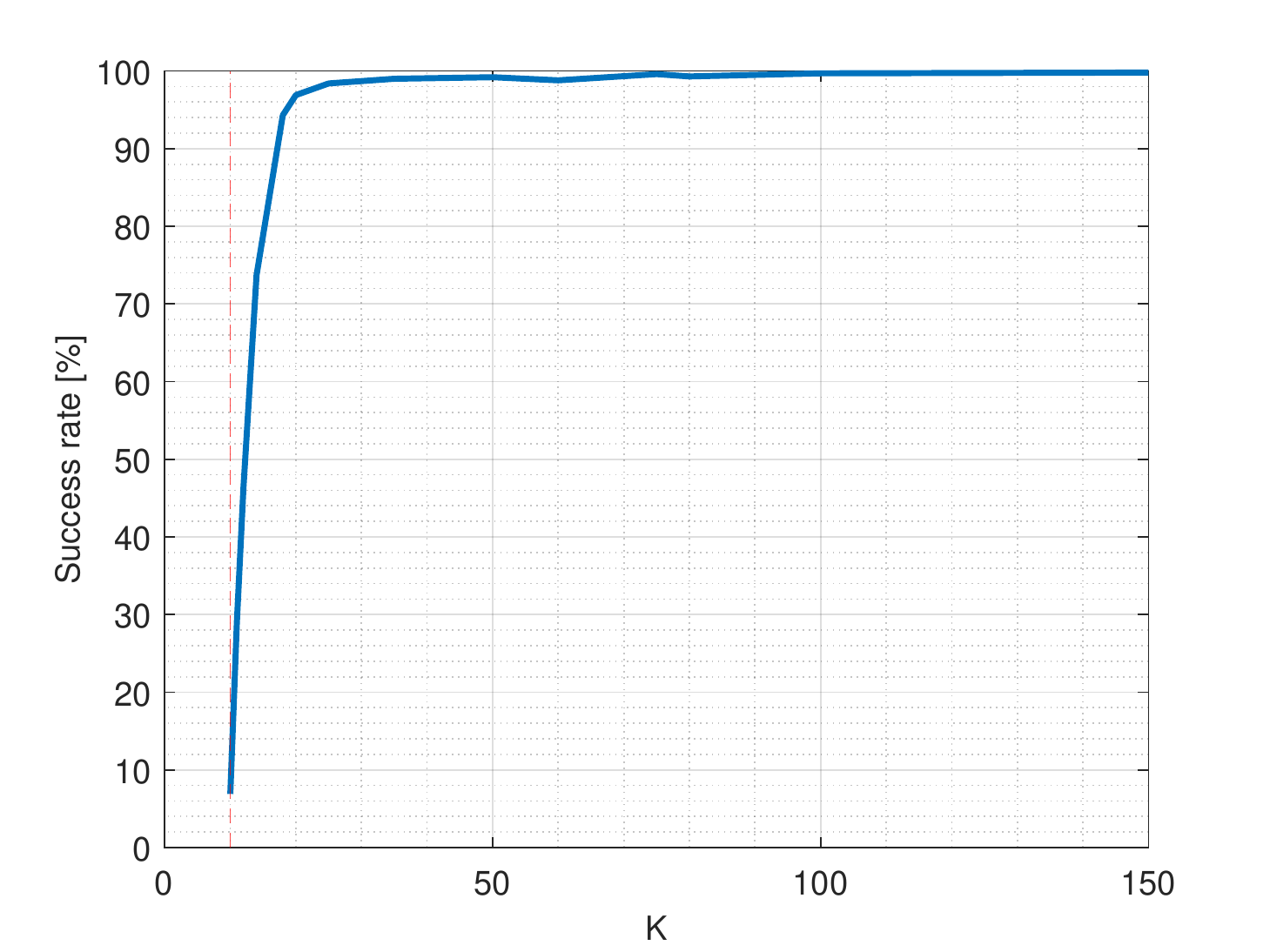}
		\caption{Trispectrum, $N=10$}
	\end{subfigure}
	\caption{\label{fig:sensing} The success rate of recovering a signal $x\in\R^N$ from $K$ random linear measurements of its bispectrum and its trispectrum. {The red vertical line specifies $N$.} As can be seen, for $K$ slightly larger than the signal's dimension, signal recovery is possible.}
\end{figure}

\paragraph{Experiment 1. Recovery from $K$ random linear  measurements.}
In this experiment, we examined the success rate of signal recovery from $K$ random  linear measurements of the bispectrum~\eqref{eq:bispectrum} and the trispectrum~\eqref{eq:trispectrum}, for different values of $K$.
In particular, each entry of the sensing matrix $A\in\R^{K\times N^q}$ was drawn from an  i.i.d.\ Gaussian distribution with zero mean and variance~1.
Figure~\ref{fig:sensing}  reports the success rate for the bispectrum 
and the trispectrum 
as a function of $K$.
As can be seen, the signal can be recovered from a few random linear measurements
 of both the bispectrum and trispectrum. Notably, the success rate is far from zero when $K$ is only slightly larger than $N$, providing a numerical support to our theoretical results.

\begin{figure}
	\begin{subfigure}[ht]{0.45\columnwidth}
		\centering
		\includegraphics[width=\columnwidth]{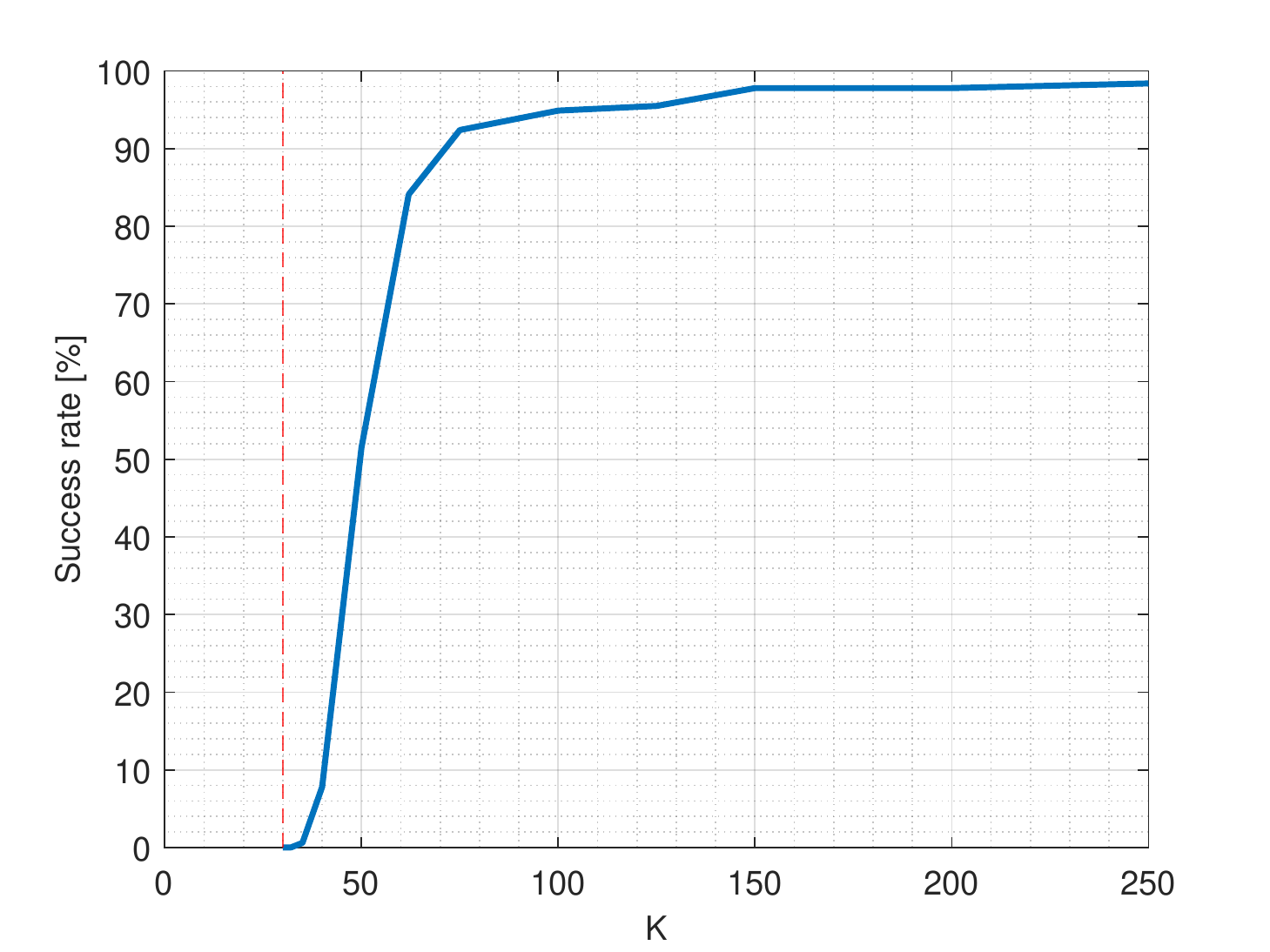}
		\caption{Bispectrum, $N=30$}
	\end{subfigure}
	\hfill
	\begin{subfigure}[ht]{0.45\columnwidth}
		\centering
		\includegraphics[width=\columnwidth]{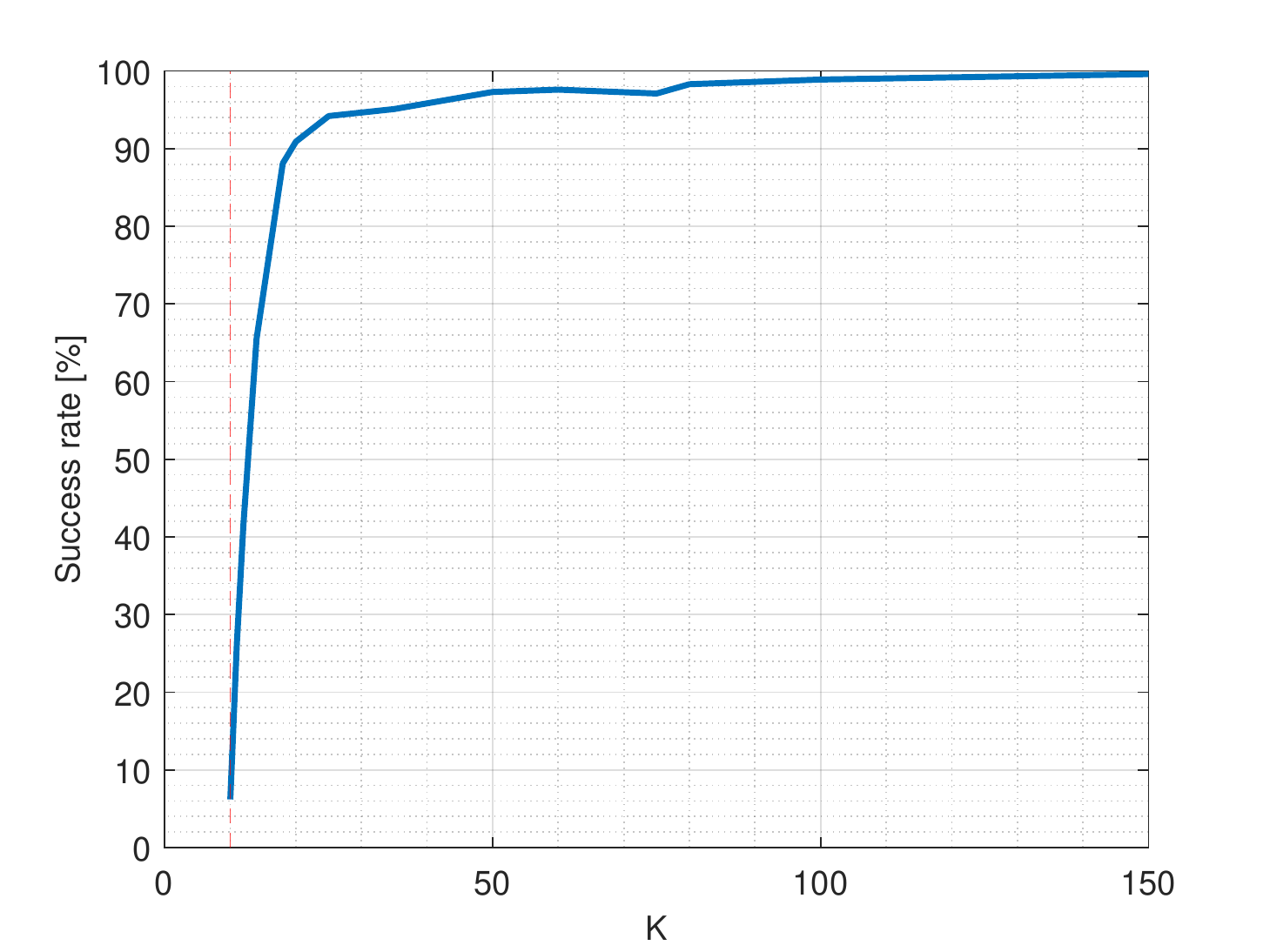}
		\caption{Trispectrum, $N=10$}
	\end{subfigure}
	\caption{\label{fig:sensing_structured} The success rate of recovering a signal $x\in\R^N$ from $K$ linear measurements; each measurement is a bispectrum (left panel) or a trispectrum (right panel) of a random signal. The red vertical line specifies $N$.
	Evidently, the results are only slightly worse than the experiment depicted in Figure~\ref{fig:sensing}.
}
\end{figure}

\paragraph{Experiment 2. Recovery from $K$ structured linear measurements.}
{The second experiment, whose results are presented in Figure~\ref{fig:sensing_structured}, examines the success rate  from $K$ linear measurements,
where each measurement is a bispectrum (left panel) or a trispectrum (right panel) of  a random signal.
Namely, each raw of the matrix $A$~\eqref{eq:measurement}  is a bispectrum  or a trispectrum of a random signal.
 The entries of the random signals were drawn  i.i.d.\ from a Gaussian distribution with zero mean and variance~1.
Therefore, in this case the measurement operator is structured, and resembles the  setup of the ultra-short pulse characterization using multi-mode fibers application  described in Section~\ref{sec:motivation}.
The results are only slightly worse than the case of random sensing vectors (Figure~\ref{fig:sensing}).
}

\begin{figure}
	\begin{subfigure}[ht]{0.45\columnwidth}
		\centering
		\includegraphics[width=\columnwidth]{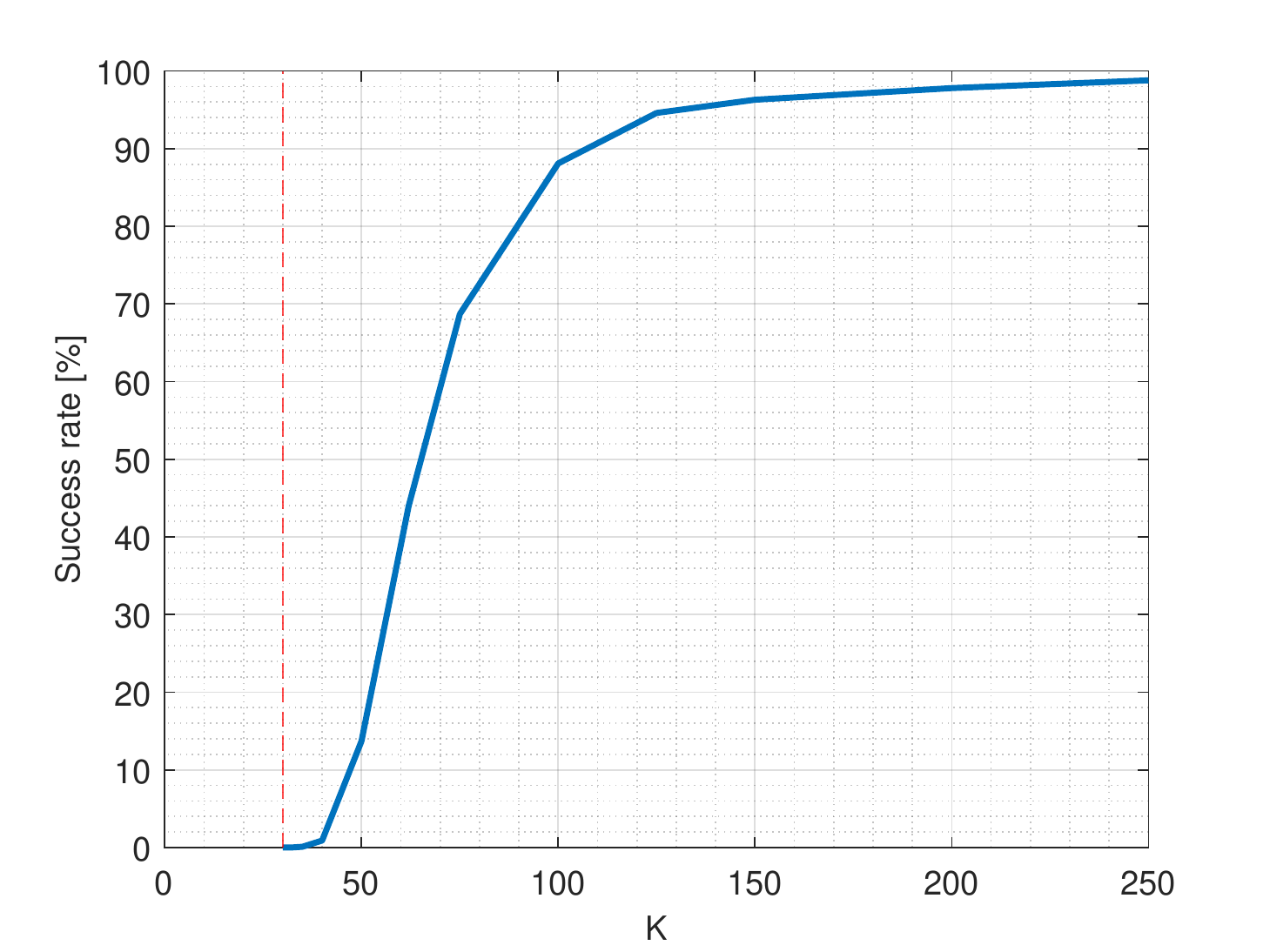}
		\caption{Bispectrum, $N=30$ }
	\end{subfigure}
	\hfill
	\begin{subfigure}[ht]{0.45\columnwidth}
		\centering
		\includegraphics[width=\columnwidth]{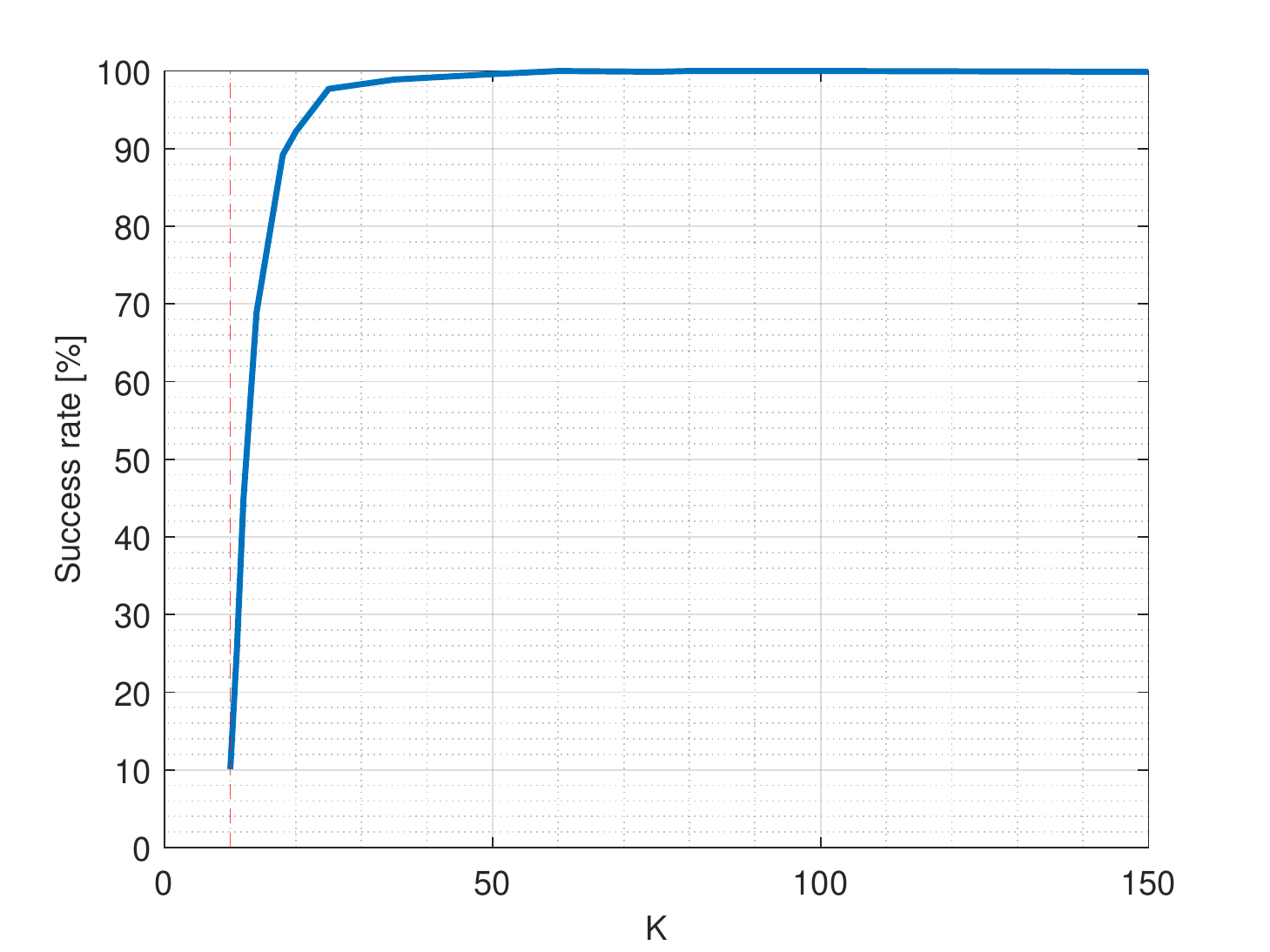}
		\caption{Trispectrum, $N=10$ }
	\end{subfigure}
	\caption{\label{fig:sampling} The success rate of recovering a signal $x\in\R^N$ from $K$ random samples of its bispectrum and its trispectrum. {The red vertical line specifies $N$.} Clearly, the success rate grows quickly for~$K$ slightly larger than~$N$, but not as fast as in Figure~\ref{fig:sensing}.}
\end{figure}

\paragraph{Experiment 3. Recovery from $K$ random samples.}
In our  second numerical experiment, we examined signal recovery from $K$ random samples of its bispectrum and  its trispectrum.
The~$K$ samples were drawn from a uniform distribution over all possible sets of size~$K$.
Figure~\ref{fig:sampling} illustrates the success rate of signal recovery from random samples  as a function of $K$.
Remarkably, for both the bispectrum and the trispectrum, we see a significant success rate for $K\ll N^{q-1}$. This
indicates that recovery is possible from a few samples of the bispectrum or the trispectrum, perhaps merely ${O}(N)$ samples.
Nevertheless, the success rates in Figure~\ref{fig:sampling} are poorer than those reported in Figure~\ref{fig:sensing}, indicating that  it is harder to recover a signal from random samples of its high-order spectra than from {random linear measurements of its high-order spectra.}

\section{Future studies} \label{sec:future}
In this paper, we have shown that one can identify a signal, up to {symmetries},
from ${O}(N)$ (generic) linear measurements of its high-order spectra (e.g., bispectrum, trispectrum).
This result has direct implications to imaging applications, such as ultra-short pulse characterization and single-particle reconstruction.
The proof is based on tools from the field of algebraic geometry. Our numerical experiments also indicate that a signal can be  recovered from a few random samples of its bispectrum or trispectrum. However, unfortunately, our proof cannot be extended to the latter case since the notion of generic measurement, as it is understood in the field of algebraic geometry, cannot be used for a finite set of possible measurement operators.
Nevertheless, we hope to fill this theoretical gap  in a future study using tools from combinatorics and probability. We formulate the following conjecture.
\begin{conj}[Random sampling of high-order spectra]
{The orbit of a} generic signal $x\in\C^N$ can be recovered from ${O}(N)$ random samples of its high-order spectra with high probability.
\end{conj}

As mentioned in Section~\ref{sec:cryoEM}, a prime motivation of this work is single-particle reconstruction using cryo-EM.
From a mathematical perspective, the cryo-EM problem~\eqref{eq:cryo_em} is a special case of the multi-reference alignment model. This model entails estimating a signal $x$ from $N$ {realizations $y_1,\ldots,y_N$ of a random variable $y$ whose distribution is characterized by }
\begin{equation} \label{eq:mra}
	y = T (g \circ x) + \varepsilon, \qquad x\in\Xi, \quad g\in G,
\end{equation}
where $T$ is a  known {(deterministic)} linear operator, $g$ is a random element  of some compact group~$G$, acting on a vector space $x\in \Xi$, and $\varepsilon$ is a noise term~\cite{bandeira2020non,bendory2020single}.
Remarkably, it was shown that the method of moments---a classical inference technique---achieves the optimal estimation rate of the multi-reference alignment model when the noise level is much larger than the signal~\cite{perry2019sample,bandeira2017estimation,abbe2018multireference,abbe2018estimation} {(in the finite-dimensional case~\cite{romanov2021multi})}.

Recall that the $q$-th statistical moment is defined as $\E\{y^{\otimes q}\}$, where the expectation is taken against the distribution of the group elements over $G$ and the noise,  $y^{\otimes q}$ is a tensor with $N^q$ entries, and the entry indexed
by $n=(n_1,\ldots,n_q)\in\Z_N^q$ is given by $\prod_{i=1}^qy[n_i]$.
Notably, in the multi-reference alignment model~\eqref{eq:mra}, {$y$}  depends linearly in~$x$, and therefore the moments are polynomials of the signal $x$.
In particular,  often times the statistical moments coincide with  high-order spectra, e.g.,~\cite{bendory2017bispectrum,bandeira2017estimation}.
We believe that under the multi-reference alignment model, our proof technique can be generalized to the identification of a signal from a few linear measurements of the high-order statistical moments of $y$.
We conjecture the following generalization of Theorem~\ref{thm.informal}.
\begin{conj}[Mutli-reference alignment] \label{conj:mra}
Suppose that the  q-th moment of {$y$~\eqref{eq:mra}} determines $x$ uniquely (possibly, up to some intrinsic symmetries).
Then, $x$ is determined uniquely (up to the same intrinsic symmetries) from  ${O}(N)$ random linear measurements or {random} samples of the $q$-th moment of~$y$.
\end{conj}

{
\begin{remark}
Recently, Conjecture~\ref{conj:mra} was  partly verified for  the dihedral multi-reference alignment model~\cite{bendory2021dihedral}.
\end{remark}
}

\section*{Acknowledgment}
We are grateful to Hui Cao and Yaron Bromberg for introducing us the fascinating application  of ultra-short pulse characterization using multi-mode fibers, which motivated this study.
{The authors also thank the reviewers for
	their insightful comments.}
T.B.  is partially supported by the NSF-BSF award 2019752, and by the Zimin Institute for Engineering Solutions Advancing Better Lives.  D.E. is supported by Simons Collaboration grant 708560. S.K. is supported by the Yitzhak and Chaya Weinstein Research Institute for Signal Processing.

\bibliographystyle{plain}

\appendix

\section{Required algebraic geometry definitions and results}
\label{sec:appendix}
Let $\K$ be a field (specifically, $\K = \R$ or $\K = \C$).
A subset of $\K^N$  which is the locus of zeros of a collection of polynomials in
$\K[x_1, . . . , x_N]$ is called an (affine) algebraic set. The
{\em Zariski topology} on~$\K^N$ is the topology whose closed sets
are algebraic subsets. (Note that the empty set and~$\K^N$ are algebraic
sets,  and arbitrary
intersections of algebraic sets are algebraic, so this defines a topology.)
A Zariski closed set is also closed in the Euclidean topology. The complement of an algebraic set is a Zariski open set. A non-empty Zariski open set is open and dense in the Euclidean topology, and its complement has Euclidean dimension strictly less than $N$.

An algebraic set $X$ is {\em irreducible} if it cannot be expressed as the union
of algebraic subsets $X_1, X_2$, with $X_1, X_2$ not empty or equal to $X$. An irreducible algebraic set is called an algebraic variety. Any algebraic set is the union of a finite number of irreducible algebraic sets. An algebraic subset~$Y$, which is a subset of an algebraic set $X$, is  called an algebraic subset
of $X$. When $Y$ is a variety (i.e., irreducible), then $Y$ is called a subvariety of $X$. The algebraic subsets of an algebraic set $X$ define a topology on $X$, which we also call the Zariski topology on
$X$. We say that a {\em generic point} of an algebraic variety $X$ has a certain property if there is a non-empty Zariski open set of points having this property.

A polynomial mapping $f \colon X \to Y$ of affine algebraic varieties is called {\em birational} if it is an isomorphism on a Zariski dense open set. More
generally, we say that $f$ is birational onto its image if the mapping
$X \to \overline{f(X)} \subset Y$ is birational. In this case, the polynomial
mapping is injective on a dense open subset of $X$ and we say that $f$
is generically injective.
More specifically, if the polynomial mapping $f$
corresponds a collection of measurements
on the vectors in
$X$, and if $f$ is birational onto its image, then we say that the {\em generic vector} can be recovered from the measurements $f$.

If $G$ is a finite group acting linearly on $\K^N$, then a classical result in invariant theory states that the set of
$G$-orbits $\K^N/G$  is an affine algebraic variety. For a reference,
see \cite[Theorem 3.5]{DolgachevInvariantTheory}. Note that $\K^N/G$ will be embedded in $\K^M$ for some $M \geq N$. For example,  if $G = \{\pm 1\}$
acting on $\C^2$ by $(-1)(a,b) = (-a,-b)$, then the quotient
$\C^2/G$ is the subvariety of $\C^3$ defined by the equation $Y^2-XZ =0$, where
$X,Y,Z$ correspond to the $\{\pm 1\}$ invariant functions $x^2, xy,y^2$ on $\C^2$.
More generally, if $X$ is an algebraic subset of $\K^N$ which is invariant
under the action of the group $G$, then $X/G$ is an algebraic subset of $\K^N/G$.
The quotient $X/G$ is characterized by the property that if $f\colon X \to Y$
is a polynomial map which is constant on $G$ orbits (i.e., $f(gx) = f(x)$ for
any $g \in G$), then $f$ factors through a polynomial map $X/G \to Y$.

\end{document}